%% file: version_04.tex
\documentclass[conference]{IEEEtran}

\ifCLASSINFOpdf
\else
\fi

\usepackage{amssymb}
\usepackage{tikz}

\usetikzlibrary{shapes}

\def\x{{\mathbf x}}

\input{packages}

\begin{document}
\title{Decentralized Detection in Energy Harvesting\\Wireless Sensor Networks}

\author{\IEEEauthorblockN{Alla Tarighati\IEEEauthorrefmark{1},
James Gross\IEEEauthorrefmark{2} and
Joakim Jald{\'e}n\IEEEauthorrefmark{1}}
\IEEEauthorblockA{\IEEEauthorrefmark{1}ACCESS Linnaeus Centre, Department of Signal Processing, KTH Royal Institute of Technology, Stockholm, Sweden\\
Email: \{allat,\,jalden\}@kth.se}
\IEEEauthorblockA{\IEEEauthorrefmark{2}ACCESS Linnaeus Centre, Department of Communication Theory, KTH Royal Institute of Technology, Stockholm, Sweden\\
Email: james.gross@ee.kth.se}
}


\maketitle

\begin{abstract}
We consider a decentralized hypothesis testing problem in which several peripheral energy harvesting sensors are arranged in parallel. Each sensor makes a noisy observation of a time varying phenomenon, and sends a message about the present hypothesis towards a fusion center at each time instance $t$. The fusion center, using the aggregate of the received messages during the time instance $t$, makes a decision about the state of the present hypothesis. We assume that each sensor is an energy harvesting device and is capable of harvesting all the energy it needs to communicate from its environment. Our contribution is to formulate and analyze the decentralized detection problem when the energy harvesting sensors are allowed to form a long term energy usage policy. Our analysis is based on a queuing-theoretic model for the battery. Then, by using numerical simulations, we show how the resulting performance differs from the energy-unconstrained case.
\end{abstract}


%
\IEEEpeerreviewmaketitle

\section{Introduction}\label{sec:intro}
Decentralized signal processing systems have received significant interests during the past decades. Because of their use in wireless sensor networks this interest has been renewed during the past years, see~\cite{Cham03,Veer12} and references therein.
In wireless sensor networks a large number of simple sensors are used. Each sensor is equipped with a small battery with limited lifetime, which in effect limits the lifetime of the sensor network. When the battery of a sensor runs out it will usually not be replaced and the sensor dies. The overall sensor network dies if a sufficient number of sensors die. To increase the lifetime of the battery-powered sensors, many solutions have been proposed~\cite{Sha10}, e.g., reducing the number of transmission bits or choosing the best modulation strategy. However, in all of these methods the sensors eventually die. To overcome this problem, an alternative is to use energy harvesting sensors that acquire their energy needed from nature or man-made sources~\cite{Ulu15,Sud11}.

Energy harvesting technology in wireless sensor networks promises a self-sustainable, maintenance free and perpetually communicating system with a lifetime that is not limited by the lifetime of individual sensors' batteries~\cite{Gun14}. While using energy harvesting devices makes it possible to deploy wireless sensor networks in hard-to-reach places, and provides potentially perpetual operation, it poses new challenges relating to the management of harvested energy because energy sources are usually sporadic and limited. For example, the source of energy might be such that energy can not be harvested all the times while we may want to use the sensors continually.

In this paper we address the problem of decentralized detection in a network of energy harvesting sensors arranged in parallel. At each time instance $t=1,2,\ldots$ the sensors send a message towards the fusion center (FC) about the perceived state of a time dependent hypothesis $H_t$ which is drawn from a binary set $\mathcal{H}\triangleq\{0,1\}$. For tractability, we assume that the noisy observations at the sensors are independent and identically distributed (iid), conditioned on the true hypothesis $H_t$. The sensors communicate with the FC through a parallel network using energy asymmetric on-off keying (OOK) where a positive message can be sent at the cost of one unit of energy, and a negative message can be conveyed though a non-transmission at no cost in energy. The sensors are allowed to use a long term energy conservation policy managed together with an internal battery.

The problem is structurally similar to the classical decentralized binary hypothesis testing problem over parallel one bit channels, and we consider the optimization of network performance in terms of the Bhattacharyya distance~\cite{Kai67} between the two hypotheses at the input of the FC~\cite{Lon90,Alla14}. Our first contribution is to show how the Bhattacharyya distance depends jointly on the sensor transmission rule and the battery depletion probability. We then illustrate the usefulness of this result by showing how to optimize the sensor transmission rule under the assumption of an unlimited-capacity battery, where we model the battery state using a birth-death process and obtain the steady state depletion probability. We then numerically compare the performance of a network of energy harvesting sensors and the performance of a network of optimally designed sensors for the \emph{unconstrained} case, where energy is always available at the sensors, in terms of Bhattacharyya distance and error probability.

The outline of this paper is as follows: In Section \ref{sec:two} we define the system model and formulate the problem. In Section \ref{sec:three} we analyze the performance of an infinite capacity energy harvesting sensor by modeling its battery as a birth-death process. In Section \ref{sec:four} we illustrate the benefits of the results by numerical simulations, and in Section \ref{sec:con} we conclude the paper.

\section{Preliminaries}\label{sec:two}
\begin{figure}
\input{topology.tex}
\caption{Decentralized detection in energy harvesting parallel network.}
\label{fig:topol}
\end{figure}
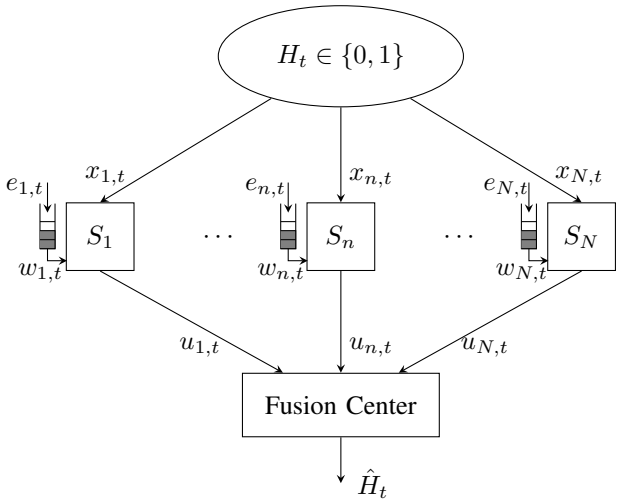
We consider a problem of decentralized hypothesis testing in which several peripheral nodes are arranged in parallel according to Fig.~\ref{fig:topol}. During time interval $t$, accounting for the time passed during $[t,t+1)$, sensor $S_n$, $n=1,\ldots,N$ makes an observation $x_{n,t}$ of the phenomenon $H_t\in\{0,1\}$ and sends a message $u_{n,t}$ towards the FC. Using the aggregate received messages $\underline{u}_t\triangleq(u_{1,t},\ldots,u_{N,t})$ the FC makes a decision ${\hat{H}}_t$ about the state of the present hypothesis at that time interval. We assume that the present hypothesis $H_t$ changes over time, while it is fixed during each time interval. There is therefore no value in the FC aggregating received messages over time to make a more reliable decision. Sensor $S_n$ consumes $w_{n,t} = w_{n,t}(u_t)$ packets of energy to send the message $u_{n,t}$ towards the FC. After sending a message, it harvests $e_{n,t}$ packets of energy from its environment and saves in its energy buffer (if possible).

We assume that $e_{n,t} \in \{0,1\}$, and energy packets arrive stochastically according to a stationary and ergodic process during time interval $t$ \cite{Ulu15}. In other words, during a time interval sensor $S_n$ is capable of harvesting at most one packet of energy.
We further assume $e_{n,t}$ is drawn from the set $\{1,0\}$, with probabilities $p_{e}, 1-p_e$ and independently for each $n$ and $t$. Concretely, sensor $S_n$ harvests a packet of energy with probability $p_{e}$ during time interval $t$ and does not with probability $1-p_{e}$. Let all the sensors have the same energy buffer (battery) size $b_{\max}$. For sensor $S_n$, the amount of available energy in the battery at the transmission time $t+1$ can not exceed $b_{\max}$ and is equal to \cite{Sha10}
\begin{equation}b_{n,t+1}=\min\big\{b_{n,t}-w_{n,t}+e_{n,t},b_{\max}\big\},\label{eq:battery}\end{equation}
where $b_{n,t}$ is the amount of energy available at sensor $S_n$ at transmission time $t$. While our initial results regarding the Bhattacharyya distance hold for this general battery model, we will in this paper assume the energy storage buffer is infinite, i.e., the battery capacity is $b_{\max}=\infty$, in order to more easily illustrate the consequences of our results. The infinite battery assumption is also a good approximation in practice for commercially available batteries, see \cite{Sha10}.

As shown in Fig.~\ref{fig:topol}, the sensors communicate with the FC through one-way parallel access channels using energy asymmetric on-off keying (OOK), where a positive message is sent at the cost of one packet of energy and a negative message is conveyed through a non-transmission at no energetic cost. It was shown in \cite{Li11} that under Rayleigh fading scenario OOK is the most energy efficient modulation scheme, though here we do not consider fading channels between the sensors and the FC. In this paper, a positive message and a negative message sent by sensor $S_n$ at time $t$ are labeled by $u_{n,t}=1$ and $u_{n,t}=0$, respectively, with the corresponding energy costs
\begin{equation*}
w_{n,t} =  w_{n,t}(u_t) = \begin{cases}
           1 & u_{n,t}=1\,,\\
		0 & u_{n,t}=0 \,. \end{cases}
\end{equation*} 

For this binary hypothesis testing problem, we assume that at each time interval $t$ the phenomenon $H_t$ is modeled as a random variable drawn from the set $\{0,1\}$ with a-prior probabilities $\pi_0$ and $\pi_1$, respectively. We also assume that observations $x_{n,t}$ at the sensors are continuous and iid over time and space, i.e., $x_{n,t}$ is viewed as an independent realization of a common random variable $X$ over some observation space $\mathcal{X}$ with a conditional probability distribution $f_{X\vert H_t}(x_{n,t}\vert h_t)$. Sensor $S_n$ is a decision maker which maps its observation $x_{n,t}$ to the output message $u_{n,t}$. Ideally, when there is no energy constraint, the sensor uses a (likelihood ratio) threshold test and sends $u_{n,t}=1$ when 
its observation $x_{n,t}$ is above a threshold $\Theta_n$, and $u_{n,t}=0$ otherwise \cite{Varsh96}. However, in an energy harvesting sensor network the action of each sensor is also limited by the battery charge $b_{n,t}$ at each transmission time $t$. Concretely, we will assume that sensor $S_n$ at transmission time $t$ sends a message $u_{n,t}=1$ towards the FC at the cost of one unit of energy if its observation $x_{n,t}$ is above a threshold $\Theta_n$ and its battery is not empty, $b_{n,t}>0$, i.e.,
\begin{equation}
u_{n,t}=  \begin{cases}
           1 & x_{n,t}\geq \Theta_n\,,\, b_{n,t}>0\,,\\
		0 & \text{Otherwise}\,. \end{cases}
\label{eq:test}\end{equation} 

Our goal is to find optimum thresholds $\Theta_1,\ldots,\Theta_N$ which maximize the total Bhattacharyya distance at the input of the FC at every time $t$. 
The total Bhattacharyya distance at the FC at time $t$ is given by
\begin{equation}\begin{split}
\mathcal{B}_{\mathrm{Tot},t}& \triangleq -\log\sum_{\underline{u}_t\in \{0,1\}^N}\sqrt{P_{\underline{U}\vert H_t}(\underline{u}_t\vert 0)P_{\underline{U}\vert H_t}(\underline{u}_t\vert 1)}
\label{eq:BD1}\end{split}\end{equation}
where $P_{\underline{U}\vert H_t}(\underline{u}_t\vert {h}_t)$ is the conditional PMF associated with the aggregate message vector $\underline{U}\triangleq(U_1,\ldots,U_N)$ that models the randomness of the message vector $\underline{u}_t \triangleq (u_{1,t},\ldots,u_{N,t})$, and is because of the independence of observations given by
\begin{equation*}
P_{\underline{U}\vert H_t}\left(\underline{u}_t\vert h_t \right)=\prod_{n=1}^N P_{U_n\vert H_t}\left(u_{n,t}\vert h_t \right) .
\end{equation*}
Due to the independence of the observations, the total Bhattacharyya distance at the FC becomes 
\begin{equation}\begin{split}
\mathcal{B}_{\mathrm{Tot},t}=\sum_{n=1}^N \mathcal{B}_{n,t}\,,
\label{eq:BD11}\end{split}\end{equation}
where 
\begin{equation}\begin{split}
\mathcal{B}_{n,t}=-\log\sum_{{u}_{n,t \in \{0,1\}}}\sqrt{P_{{U}_n\vert H_t}({u}_{n,t}\vert 0)P_{{U}_n\vert H_t}({u}_{n,t}\vert 1)}
\label{eq:BD12}\end{split}\end{equation}
 is the Bhattacharyya distance at the FC resulting from sensor $S_n$. The motivation of using the Bhattacharyya distance as performance metric is that using the probability of error at the FC makes the design procedure intractable. Also when the FC is a maximum a-posteriori (MAP) detector, it is possible to upperbound the minimum probability of error at the FC using the Bhattacharyya distance according to \cite{Lon90}
\[P_{\mathrm{E},t}\leq \sqrt{\pi_0\pi_1}\,e^{-\mathcal{B}_{\mathrm{Tot},t}}.\]

According to \eqref{eq:BD11} the total Bhattacharyya distance at the FC is the summation of the delivered Bhattacharyya distances from individual sensors. This greatly simplifies the problem of jointly designing the sensor decisions to the problem of designing a set of individual and identical sensor decision rules (thresholds). Therefore, in what follows, we drop the subscript $n$ and focus on finding optimum threshold $\Theta^\star$ which maximizes the Bhattacharyya distance of a single sensor $S$.

\section{Analyzing an Energy Harvesting Sensor}\label{sec:three}
In this section we will be considering the Bhattacharyya distance of a single energy harvesting sensor. We will show how this can be formulated in terms of its threshold $\Theta$ and the depletion probability of its battery. Let $B_t$ be a random variable corresponding to the battery charge $b_t$. By \eqref{eq:test}, the conditional mass probabilities of the sensor decision are 
\begin{align}
P_{U\vert H_t}\left(u_t=1\vert h_t\right)&=\Pr\left( X\geq \Theta \cap B_t>0\vert H_t = h_t  \right) \label{eq:Pu1}\\
&=\Pr\left( X\geq \Theta\vert H_t=h_t \right)\Pr\left( B_t>0 \right) \nonumber \\
&=\Pr\left( X\geq \Theta\vert H_t=h_t \right)\left[1-\Pr\left( B_t=0 \right)\right] \nonumber,
\end{align}
since the available energy $B_t$ at transmission time $t$ is independent of the true hypothesis $H_t$ and the observation $X$ at the same time. 

Under the assumptions that the energy harvesting probability is uncorrelated in time and has a fixed probability $p_e$, and the observations at the sensor are iid in time, the battery has a Markovian behavior in the sense that its state at the transmission time $t+1$ (i.e., $B_{t+1}$), conditioned on $E_t$ and $W_t$, only depends on its state at transmission time $t$  (i.e., $B_t$) and not the sequence of previous states, $\{B_{t'} \}_{t'=0}^{t-1}$, where $E_t$ and $W_t$ are the random variables corresponding to $e_t$ and $w_t$, respectively. This can also be seen from \eqref{eq:battery}.  Under the Markovian assumption and the time invariant transmission policy, the battery state will have a stationary distribution, which allows us to consider the steady state performance of an energy harvesting sensor. To this end, we drop the subscript $t$ from \eqref{eq:Pu1}, and define $p_{k}$ for $k=0,1,\ldots$ as the steady state probability that the battery is in state $k$. In other words, $p_{k}$ is the long-term fraction of time that the battery charge at an arbitrary transmission time is $k$ packets of energy, and accordingly $p_{0} = \Pr\left( B_t=0 \right)$ is the steady-state depletion probability of the battery. Then \eqref{eq:Pu1} reduces to
\begin{equation}\begin{split}
P_{U\vert H}\left(u=1\vert h\right)=P_h(X; \Theta)\left(1-p_{0} \right),
\label{eq:Pu2}\end{split}\end{equation}
where $P_h(X; \Theta)\triangleq\Pr\left( X\geq \Theta\vert H_t=h \right)$. Using \eqref{eq:Pu2}, the Bhattacharyya distance of a single energy harvesting sensor in steady state can be expressed as
\begin{align}
\mathcal{B}=-&\log \bigg[\left(1-p_{0}\right)\sqrt{P_0(X; \Theta)\,P_1(X; \Theta)}\,+ \label{eq:BD2}\\
&\sqrt{\big[1-P_0(X; \Theta)\left(1-p_{0} \right) \big]\!\big[1-P_1(X; \Theta)\left(1-p_{0} \right) \big]}\bigg] \nonumber.
\end{align}
We observe from \eqref{eq:BD2} that the Bhattacharyya distance of an energy harvesting sensor not only depends on the choice of the threshold $\Theta$, but it also depends on the depletion probability $p_0$. The depletion probability of the battery is itself a function of energy features, battery size, and the threshold $\Theta$ itself. 
\begin{figure}
\input{birthdeath.tex}
\caption{A birth-death process.}
\label{fig:bd}
\end{figure}
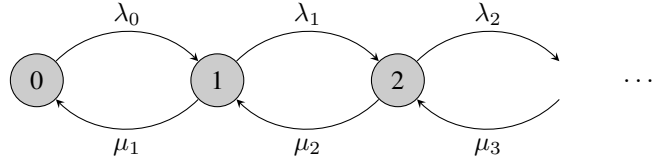

We will next model the battery state as a birth-death process and find a closed-form expression for the depletion probability of the battery. This gives us a tool for finding an optimum threshold $\Theta^\star$ of an energy harvesting sensor. To this end, let $S$ be an energy harvesting sensor with an unlimited-capacity battery. In this case there always exists space for harvested energy. Let at transmission time $t$ the battery be at state $k$, $k>0$. Then at transmission time $t+1$ its state will be either $k-1$, $k$, or $k+1$. This is due to the fact that during each time interval at most one packet of energy can be harvested or be consumed. Note that if the battery at transmission time $t$ is at state zero, $b_t=0$, its state at time $t+1$ would be either zero or one, since the battery charge can not be negative. The transition probabilities in steady state can be written as
\begin{align}
p_0&=p_{0,0}p_0+p_{1,0}p_1, \nonumber \\
p_k&=p_{k-1,k}p_{k-1}+p_{k,k}p_{k}+p_{k+1,k}p_{k+1},\quad k\geq 1,
\label{eq:SS1}\end{align}
where $p_{i,j}\triangleq\Pr\left(B_{t+1}=j\vert B_t=i\right)$ and where according to the structure of the problem we find
\[p_{0,0}=1-p_{e}\,,\quad p_{1,0}=P(X; \Theta)\left(1-p_{e}\right)\,,\quad p_{0,1}=p_{e}\,,\]
where $P(X; \Theta)\triangleq \pi_0 P_0(X; \Theta)+\pi_1 P_1(X; \Theta)$, and for $k\geq 1$
\begin{equation*}\begin{split}
p_{k,k+1}&=(1-P(X; \Theta))p_{e}\,,\\
p_{k+1,k}&=P(X; \Theta)\left(1-p_{e}\right),\\
p_{k,k}&=P(X; \Theta)p_{e}+(1-P(X; \Theta))\left(1-p_{e}\right).
\end{split}\end{equation*}
The above steady state equations represent a birth-death process \cite{Gro08}, as shown in Fig.~\ref{fig:bd}, with parameters
\begin{align*}
\lambda_0&=p_{e} \,,\\
\lambda_k&=(1-P(X; \Theta))p_{e}\,, &k\geq1,\\
\mu_k&=P(X; \Theta)(1-p_{e})  &k \geq 1.
\end{align*}

According to the flow balance \cite{Gro08}, the rate of transitions out of a given state $k$ must be equal to the rate of transitions into that state, or
\begin{align}
\lambda_0p_{0}&=\mu_1p_1 \nonumber \\
(\lambda_k+\mu_k)p_k&=\lambda_{k-1}p_{k-1}+\mu_{k+1}p_{k+1}, \quad k \geq1 .
\label{eq:bd2}\end{align}
In \eqref{eq:bd2}, the left hand side is the long term transitions out of state $k$, and the right hand side is the long term transitions into that state. Comparing \eqref{eq:SS1} and \eqref{eq:bd2}, and with slight mathematical manipulations, we can obtain desired values for $\lambda_0,\lambda_1,\ldots$ and $\mu_1,\mu_2,\ldots$. 

Modeling the state of the unlimited-capacity battery as a birth-death process, makes it possible to find a closed-form expression for its depletion probability (and Bhattacharyya distance). 
For a birth-death process, in equilibrium, the probability flows across each cut are balanced \cite{Gro08}, and we have $$\lambda_k\,p_{k}=\mu_{k+1}\,p_{{k+1}}\,,\quad k\geq 0\,.$$
This means 
that all the state probabilities $p_{k}$ can be expressed in terms of that of the state zero, $p_{0}$. By noting that the state probabilities should sum to one, i.e.,
$$\sum_{k=1}^{\infty}p_k=1$$
the depletion probability in the steady state is obtained as
$p_{0}=\frac{1}{1+\Lambda}\,,$
where \cite{Gro08} \[\Lambda\triangleq \sum_{k=1}^\infty\frac{\lambda_0\ldots\lambda_{k-1}}{\mu_{1}\ldots\mu_k}=\frac{\lambda_0}{\mu_1}\sum_{k=1}^\infty\left(\frac{\lambda_1}{\mu_1}\right)^{k-1}\,.\]

The depletion probability is zero if ${\lambda_1}\geq{\mu_1}$ (or $\Lambda=\infty$), which is equivalent to $p_{e}\geq P(X; \Theta)$. Else if ${\lambda_1}<{\mu_1}$ (or $\Lambda<\infty$), the summation above converges. The overall depletion probability is
\begin{equation}
p_{0}=\left \{
  \begin{tabular}{cc}
 $0$ & $p_{e}\geq P(X; \Theta)$\,,\\
 $1-\frac{p_{e}}{P(X; \Theta)}$ & Otherwise\,.
  \end{tabular}
\right.
\label{eq:pb0i}\end{equation}

Note that the condition $p_{e}\geq P(X; \Theta)$, which results in zero depletion probability, follows intuition in the sense that, when the probability of energy arrival $p_{e}$ is more than the probability of energy consumption (or the probability that the sensor decides to send a message $P(X; \Theta)$), the battery will asymptotically accumulate energy and will with probability one not be empty. In this situation the distributed detection problem will be the same as in the unconstrained setup.

The depletion probability in \eqref{eq:pb0i} can be used to find the Bhattacharyya distance in \eqref{eq:BD2}. From \eqref{eq:pb0i} we observe that the depletion probability, and consequently the Bhattacharyya distance, depends only on the harvesting probability $p_{e}$ and the unconstrained transmission probability $P(X; \Theta)$ though the threshold $\Theta$. Though the energy harvesting probability is assumed fixed and out of our control, we can maximize the Bhattacharyya distance by choosing an optimal threshold $\Theta^\star$. 

In the next section, we will exemplify these results by considering the optimal Bhattacharyya distance of a single energy harvesting sensor and the error probability performance of a network of such energy harvesting sensors.

\section{Numerical Results}\label{sec:four}
For the purpose of the numerical illustrations, we consider the case where each real valued observation $x_{n,t}$ is either from a Rayleigh distribution with scale parameter $\sigma_0$ or a Rician distribution with scale parameter $\sigma_1$ and non-centrality parameter $s$. We assume $\sigma_0=\sigma_1=1$, and the conditional distributions at the sensor are therefore as
\begin{equation}\begin{split}
f_{X\vert H_t}\left( x\vert 0 \right)&=xe^{-\frac{x^2}{2}}\\
f_{X\vert H_t}\left( x\vert 1\right)&=xe^{-\frac{x^2+s^2}{2}}I_0(xs)\,,
\label{eq:dist}\end{split}\end{equation}
where $I_0(z)$ is the modified Bessel function of the first kind with order zero. This observation model corresponds to an energy harvesting sensor applied to detect the presence of a known signal in Gaussian noise by received power, a relevant case for low complexity sensors in a wireless sensor networks.

\begin{figure}[t]
\centering
\includegraphics[width=0.95\columnwidth]{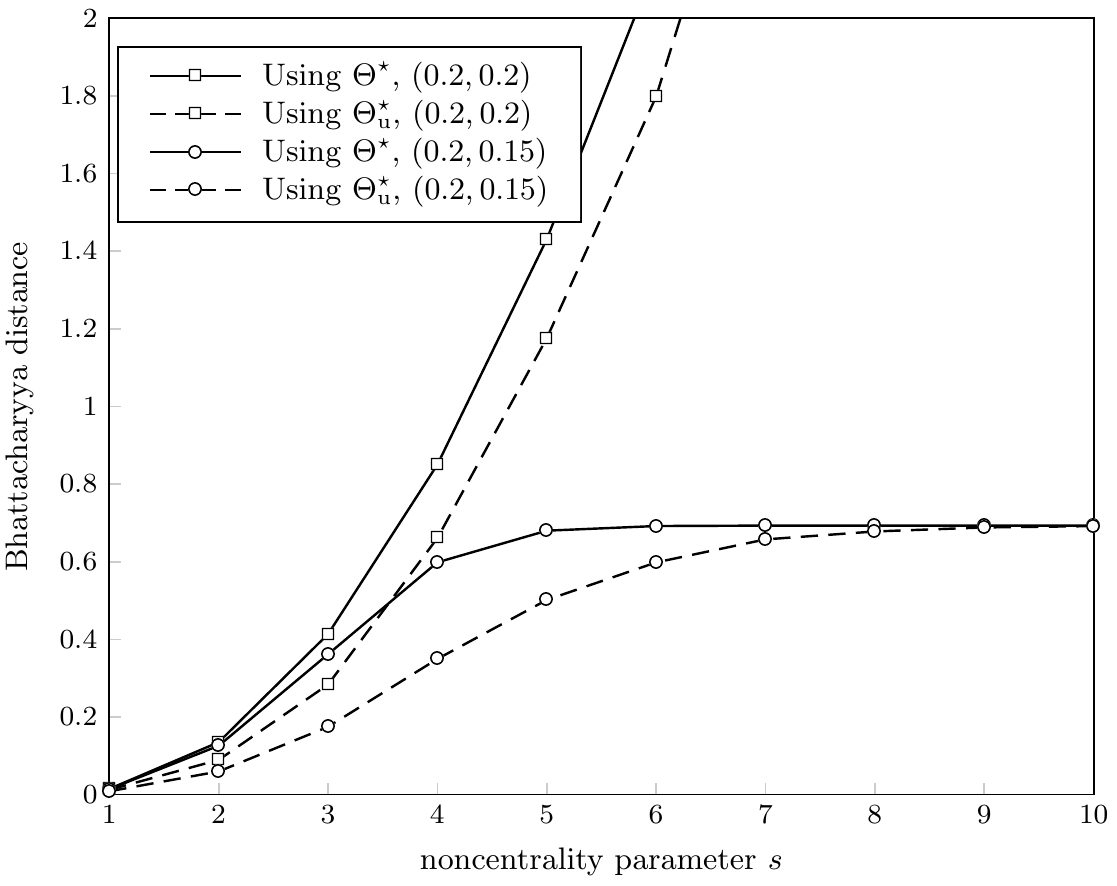}
\caption{Evolution of Bhattacharyya distance of an unlimited-capacity battery energy harvesting sensor while using unconstrained $\Theta^\star_{\mathrm{u}}$ and adapted $\Theta^\star$ thresholds and for different $(\pi_1,p_{e})$.}
\label{fig:Bhat}
\end{figure}

For energy unconstrained sensors, the optimal threshold is
\begin{equation*}\begin{split}
{\Theta}^\star_{\mathrm{u}}=\arg\max_{\Theta}\bigg\{\!\!-\log\!\bigg[&\sqrt{ P_0(X; \Theta)\,P_1(X; \Theta)}\,+\\&\sqrt{(1-P_0(X; \Theta))(1-P_1(X; \Theta))} \,\bigg]\bigg\}.
\end{split}\end{equation*}
We use this threshold as a benchmark and compare the Bhattacharyya performance of a single energy harvesting sensor applying $\Theta^\star_{\mathrm{u}}$ and a sensor that uses the energy optimal threshold $\Theta^\star$ which maximizes \eqref{eq:BD2}. Fig.~\ref{fig:Bhat} illustrates this comparison for different pairs of $(\pi_1,p_e)$. To find an optimal threshold we sweep the threshold $\Theta$ and find a threshold which maximizes the Bhattacharyya distance for both the constrained and the unconstrained cases. We observe, as expected, from Fig.~\ref{fig:Bhat} that the resulting Bhattacharyya distance using the energy optimal threshold $\Theta^\star$ outperforms that of using unconstrained threshold $\Theta^\star_{\mathrm{u}}$.

\begin{figure}[t]
\centering
\includegraphics[width=0.965\columnwidth]{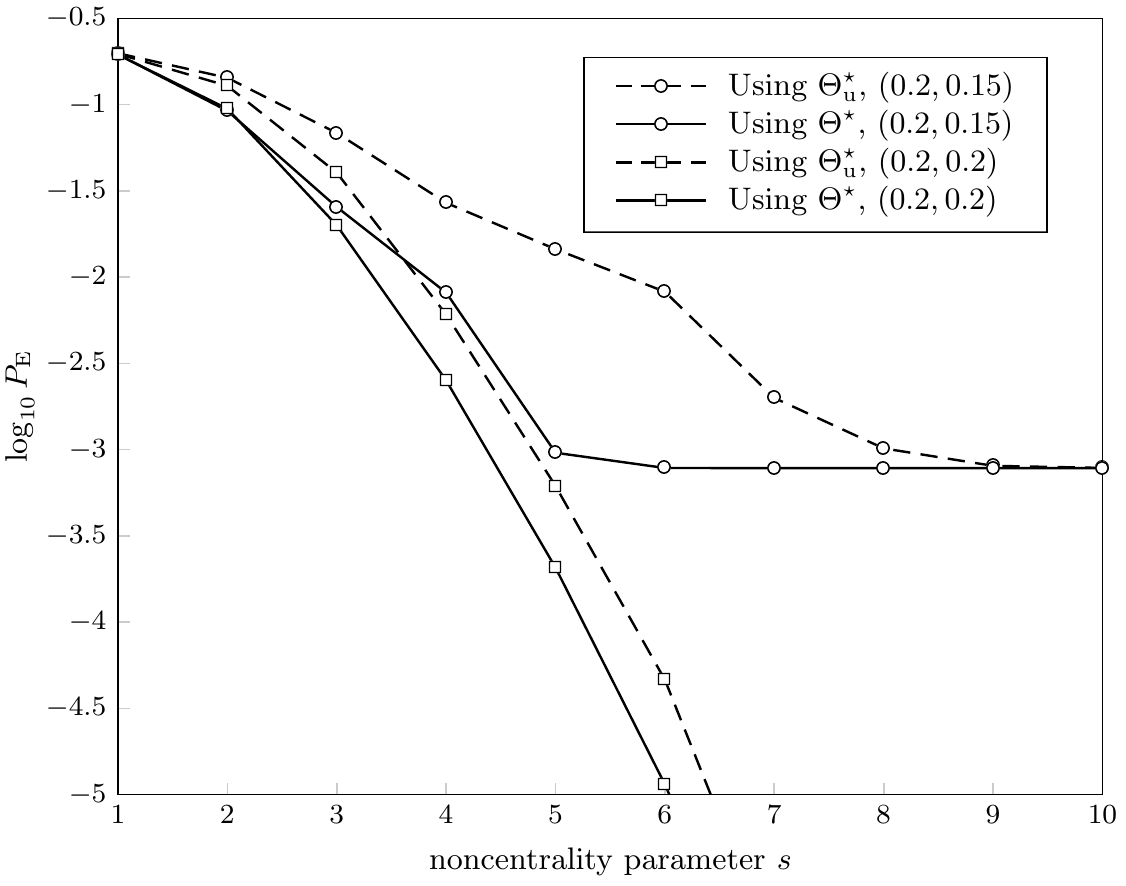}
\caption{Error probability performance of a network of $N=4$ unlimited-capacity battery energy harvesting sensors while using unconstrained $\Theta^\star_{\mathrm{u}}$ and adapted $\Theta^\star$ thresholds and for different $(\pi_1,p_{e})$.}
\label{fig:Pe}
\end{figure}

We further compare the performance of a network of $N=4$ energy harvesting sensors arranged as in Fig.~\ref{fig:topol}, and with the observation model given by \eqref{eq:dist}. By maximizing the Bhattacharyya distance of single sensors (as in \eqref{eq:BD2}) we find the optimal threshold for each sensor and then compare the error probability performance of the network with the case where the sensors use threshold $\Theta^\star_{\mathrm{u}}$. We assume that the FC using the MAP rule makes the final decision about the state of the true hypothesis at each time $t$. The error probability at time $t$ at the FC is found using
\begin{equation*}
P_{\mathrm{E},t}=1-\sum_{\underline{u}_t}\max_{h=0,1}\left\{ \pi_{h}P_{\underline{U}\vert H_t}\left(\underline{u}_t\vert h\right) \right\}.
\end{equation*}
This can be computed numerically, without the need for Monte-Carlo simulations, see \cite{Alla14}. Fig.~\ref{fig:Pe} shows the error probability performance of a network of unlimited-capacity battery sensors for different sets of $(\pi_1,p_{e_1})$. The results parallel those obtained for the Bhattacharyya distance, and show that the energy optimal threshold $\Theta^\star$ results in better error probability performance.

From the numerical results, we observe that Bhattacharyya distance (error probability), as the noncentrality parameter grows, for some cases grows (decreases) unboundedly and for some cases converges to an asymptote. This can be understood as follows: As the noncentrality parameter $s$ goes to infinity, the sensor will detect the signal without error wherever $H_t = 1$ for any reasonable choice of threshold $\Theta$, and $\pi_1$ becomes the ideal energy consumption rate. When $\pi_1$ increases it is more probable that $H_t=1$ and consequently the sensor $S$ will consume more energy to notify the FC of the presence of the signal. When $p_e < \pi_1$, the sensor will ultimately be limited by battery depletion events. When $p_e \geq \pi_1$, the sensor shows the same behavior as an unconstrained sensor, i.e., its Bhattacharyya distance grows unboundedly as the noncentrality parameter increases. The conditions under which the Bhattacharyya distance (error probability), as the noncentrality parameter (or SNR in general) increases (decreases) unboundedly will be studied extensively in \cite{Alla16-EHS-TSP}, where also the asymptotes are found under a range of different battery capacities.
\section{Conclusions}\label{sec:con}
In this paper we considered the performance a network of unlimited-capacity battery energy harvesting sensors. Using the Bhattacharyya distance, we formulated the problem of designing energy harvesting sensors in wireless sensor networks. We analyzed the performance of the sensors using a queuing-theoretic model for each sensor battery. In this paper the depletion probability and the Bhattacharyya distance performance of an unlimited-capacity battery sensor were obtained. An extended analysis can be completed by considering the performance of an arbitrary capacity battery energy harvesting sensor.\vspace{-0.4em}
\bibliographystyle{IEEEtran}
\bibliography{Ref}
\end{document}

%% file: packages.tex
\usepackage{array,booktabs}
\usepackage{algpseudocode}
\usepackage{algorithm}
\usepackage{amsfonts}
\usepackage{amsmath}
\usepackage{amssymb}
\usepackage{amsthm}
\usepackage{graphicx}
\usepackage{stmaryrd}
\usepackage{siunitx}
\usepackage{multicol}
\usepackage{tikz}
\usepackage{setspace}
\usepackage{pstricks, pst-node, pst-plot, pst-circ}
\usepackage{moredefs}
\usetikzlibrary{shapes}

\makeatletter
\newtheoremstyle{newdefinition}
  {3pt}
  {3pt}
  {}
  {1em}
  {\itshape}
  {:}
  {.5em}
  {\thmname{#1}\thmnumber{\@ifnotempty{#1}{ }#2}%
   \thmnote{ {\the\thm@notefont(\itshape#3)}}}
\makeatother
\theoremstyle{newdefinition}

\makeatletter
\newtheoremstyle{newlemma}
  {3pt}
  {3pt}
  {}
  {1em}
  {\itshape}
  {:}
  {.5em}
  {\thmname{#1}\thmnumber{\@ifnotempty{#1}{ }#2}%
   \thmnote{ {\the\thm@notefont(\itshape#3)}}}
\makeatother
\theoremstyle{newlemma}

\makeatletter
\newtheoremstyle{newtheorem}
  {3pt}
  {3pt}
  {}
  {1em}
  {\itshape}
  {:}
  {.5em}
  {\thmname{#1}\thmnumber{\@ifnotempty{#1}{ }#2}%
   \thmnote{ {\the\thm@notefont(\itshape#3)}}}
\makeatother
\theoremstyle{newtheorem}

\makeatletter
\newtheoremstyle{newproposition}
  {3pt}
  {3pt}
  {}
  {1em}
  {\itshape}
  {:}
  {.5em}
  {\thmname{#1}\thmnumber{\@ifnotempty{#1}{ }#2}%
   \thmnote{ {\the\thm@notefont(\itshape#3)}}}
\makeatother
\theoremstyle{newproposition}

\makeatletter
\newtheoremstyle{newproof}
  {3pt}
  {3pt}
  {}
  {2em}
  {\itshape}
  {:}
  {.5em}
  {\thmname{#1}\thmnumber{\@ifnotempty{#1}{ }#2}%
   \thmnote{ {\the\thm@notefont(\itshape#3)}}}
\makeatother
\theoremstyle{newproof}

\providelength{\AxesLineWidth}       
\providelength{\plotwidth}           
\providelength{\LineWidth}           
\providelength{\MarkerSize}          
\newrgbcolor{GridColor}{0.8 0.8 0.8}%

%% file: topology.tex
\begin{tikzpicture}[align=center,scale=0.8,>=stealth] 

\def\battery#1#2#3#4{
\begin{scope}[shift={#1}, rotate=#2]
\draw [-] (0,.75) -- (0,0) -- (0.25,0) -- (0.25,0.75);
\draw [-] (0,.45) -- (0.25,0.45);
\draw [-,fill=gray] (0,.3) -- (0.25,0.3) -- (0.25,0) -- (0,0);
\draw [-] (0,.15) -- (0.25,0.15);
\draw [->] (0.125,1.1) -- (0.125,0.6);
\node (e) at (-0.25,1)  {#3}; 
\draw [->] (0.125,0) -- (0.125,-0.2) -- (0.45,-0.2); \node at (0.01,-.4) {#4};
\end{scope}
}

\node (FC) at (0,0.2) [rectangle,inner sep=3mm,draw] {Fusion Center};
\node (DM1) at (-4,3) [rectangle,minimum size=0.9cm,draw] {$S_1$};
\node (DM2) at (-2,3)  {$\cdots$};
\node (DMN1) at (2,3) {$\cdots$};
\node (DMN) at (4,3) [rectangle,minimum size=0.9cm,draw] {$S_N$};
\node (DMn) at (0,3)  [rectangle,minimum size=0.9cm,draw] {$S_{n}$};
\node (PH) at (0,6) [ellipse,inner sep=3mm,draw] {$H_t\in\{0,1\}$};
\draw [->] (DM1.south) --  (FC) node [near end,left,inner sep=6pt] {$u_{1,t}$};
\draw [->] (DMn.south) -- (FC) node [near end,right,inner sep=3pt] {$u_{n,t}$};
\draw [->] (DMN.south) -- (FC) node [near end,right,inner sep=6pt] {$u_{N,t}$};
\draw [->] (FC.south) -- (0,-1.1) node [right,inner sep=6pt] {$\hat{H}_t$};
\draw [->] (PH) --(DM1.north) node [near end,left,inner sep=6pt] {$x_{1,t}$};
\draw [->] (PH) -- (DMn.north) node [near end,right,inner sep=3pt] {$x_{n,t}$};
\draw [->] (PH) -- (DMN.north) node [near end,right,inner sep=6pt] {$x_{N,t}$};
\battery{(-5,2.8)}{0}{$e_{1,t}$}{$w_{1,t}$}
\battery{(-1,2.8)}{0}{$e_{n,t}$}{$w_{n,t}$}
\battery{(3,2.8)}{0}{$e_{N,t}$}{$w_{N,t}$}
\end{tikzpicture}

%% file: birthdeath.tex
\begin{tikzpicture}[align=center,scale=0.8,>=stealth]

\pgfmathtruncatemacro{\num}{3}; 
\pgfmathtruncatemacro{\nu}{\num-1}; 
\pgfmathtruncatemacro{\numb}{\num+1}; 

\foreach \i in {1,...,\num}
{
\pgfmathtruncatemacro{\y}{\i -1};
\node[circle,draw=black,fill=white!80!black,minimum size=20] (\i) at (3*\i,0) {\y};
}

\foreach \i in {1,...,\nu}
{
\pgfmathtruncatemacro{\x}{\i +1};
\pgfmathtruncatemacro{\y}{\i -1};
\draw [->]   (\i) to[out=45,in=135] (\x);\node at (1.5+3*\i,1.1){$\lambda_\y$};
\draw [->]   (\x) to[out=-135,in=-45] (\i);\node at (1.5+3*\i,-1.1){$\mu_\i$};
}

\foreach \i in {\numb}
{
\pgfmathtruncatemacro{\z}{\i -2};
\pgfmathtruncatemacro{\y}{\i -1};
\node[circle,draw=white,minimum size=20] (\i) at (3*\i,0) {};
\draw [->]   (\y) to[out=45,in=135] (\i);\node at (1.5+3*\y,1.1){$\lambda_\z$};
\draw [->]   (\i) to[out=-135,in=-45] (\y);\node at (1.5+3*\y,-1.1){$\mu_\y$};
\node at  (1+3*\i,0) {$\ldots$};
}

\end{tikzpicture}